\documentclass[aps,prd,preprint,superscriptaddress,amsmath,amssymb,showpacs]{revtex4-1}
\usepackage{dcolumn}
\usepackage{graphicx}
\usepackage{float}
\usepackage{physics}
\usepackage[colorlinks=true,allcolors=blue]{hyperref}
\begin{document}

\title{Potential and string breaking of doubly heavy baryon at finite temperature and chemical potential
}

\author{Bo Yu}
\affiliation{School of Nuclear Science and Technology, University of South China, Hengyang 421001, China}

\author{Xi Guo}
\affiliation{School of Nuclear Science and Technology, University of South China, Hengyang 421001, China}

\author{Xun Chen}
\email{chenxunhep@qq.com}
\affiliation{School of Nuclear Science and Technology, University of South China, Hengyang 421001, China}

\author{Xiao-Hua Li}
\email{lixiaohuaphysics@126.com}
\affiliation{School of Nuclear Science and Technology, University of South China, Hengyang 421001, China}

\date{\today}

\begin{abstract}
Using gauge/gravity duality, we first study the string breaking and melting of doubly heavy baryon at a finite chemical potential and temperature in this paper. The decay mode $\rm{Q Q q \rightarrow Q q q+Q \bar{q}}$ is investigated with the presence of temperature and chemical potential in this paper. With the increase of temperature and chemical potential, string breaking takes place at a smaller potential energy and the string-breaking distance will increase slightly. It is also found that the $\rm{QQq}$ melts at small separate distance with the increase of temperature and chemical potential. Then, we compare the screening distance of $\rm{QQq}$ with $\rm{Q \bar{Q}}$ under the same conditions. Finally, we draw the melting diagram of $\rm{QQq}$ and $\rm{Q \bar{Q}}$ in the $T-\mu$ plane.
\end{abstract}

% keywords: finite chemical potential\sep holographic QCD\sep potential

\maketitle

\section{Introduction}\label{sec:01_intro}
The interquark potential is one of the crucial factors in determining the formation of baryonic bound states, and investigating the potential can facilitate a further understanding of the structure of baryons and the dynamical mechanisms of QCD.
Due to computational limitations, predicting the properties of hadrons remains a daunting challenge in QCD. Research on heavy quarks and static test charges is valuable for probing the confinement properties of QCD, as they share similar characteristics.

Lattice QCD remains the most reliable theoretical method for studying nonperturbative physics in QCD~\cite{Lang:1982tj,Hoek:1987uy,Michael:1990az,Takahashi:2002bw,Aoki:2005vt,Ratti:2005jh,Bicudo:2007xp,Luscher:2010iy}. Despite QCD at finite chemical potential can be formulated on the lattice~\cite{Hasenfratz:1983ba}; however, standard Monte Carlo techniques cannot be used at $\mu \neq$ 0. Some methods have been proposed to address this issue~\cite{Fodor:2001au,Muroya:2003qs}, but the study of lattice QCD under finite temperature and chemical potential is still challenged due to the presence of the fermion sign problem. As we know, high temperatures, densities and other extreme conditions are produced in relativistic heavy-ion collision experiments. It is essential to compare various nonperturbative methods to obtain a more comprehensive theoretical understanding of the nonperturbative physics in QCD \cite{Brambilla:2006zx,Ghiglieri:2011fhu,Andreev:2020pqy,Andreev:2021bfg,Andreev:2019cbc}.

The success of the Cornell model in describing quarkonium spectroscopy has had a profound impact on the further development of potential energy models and other approaches to nonrelativistic QCD{ \cite{Eichten:1978tg,Eichten:1979ms,Andreev:2020xor}}.
As a crucial ingredient in our understanding of strong interactions, the study of quark-antiquark potential is already sufficient through the study of a deformed $\rm{AdS_5}$ model and the Einstein-Maxwell-Dilation model, etc~\cite{Sumino:2004ht,Andreev:2006eh,Balitsky:1985iw,deForcrand:2005vv,Andreev:2006nw,He:2010bx,Colangelo:2010pe,DeWolfe:2010he,Li:2011hp,Fadafan:2011gm,Fadafan:2012qy,Cai:2012xh,Li:2012ay,Fang:2015ytf,Yang:2015aia,Zhang:2015faa,Ewerz:2016zsx,Chen:2017lsf,Arefeva:2018hyo,Chen:2018vty,Bohra:2019ebj,Bohra:2019ebj,Chen:2019rez,Zhou:2020ssi,Zhou:2021sdy,Chen:2020ath,Chen:2021gop,Gross:1998gk}.

The applicability of perturbative methods to high-energy processes and phenomena is limited by the large coupling of strong interactions in the low-energy region. The development of the holographic principle and the AdS/CFT correspondence \cite{Maldacena:1997re} provides us with a new approach to study strongly coupled nuclear matter in the low-energy region and at energy scales close to the deconfined phase transition temperature.

A holographic model developed from string theory has been proposed, which employs the Nambu-Goto action of strings at zero and finite temperature in higher-dimensional spacetime to calculate the Coulomb potential and screening effect between quarks and antiquarks\cite{Maldacena:1998im,Rey:1998ik,Rey:1998bq}. These researches provide a new approach for researching meson interactions.
A holographic baryon is considered to be a D-brane that wraps the inner subspace of the background spacetime, with a string connecting it and extending to the boundary \cite{Witten:1998xy,Gross:1998gk}.
Furthermore, Andreev proposed a string theoretical structure for the $\rm{QQq}$ system based on previous studies on the heavy quark potential in ~\cite{Andreev:2006ct,Andreev:2012mc,Andreev:2015iaa,Andreev:2015riv}, enabling us to investigate its ground state.

The basic task of this paper is to extend effective string models of holographic QCD to finite temperatures and chemical potentials.
In this model, two heavy quarks are connected by string to the baryon vertex which is a five-brane; the light quark is a tachyon field coupled to the world-sheet boundary. The rest of the paper is organized as follows: Different configurations are discussed at finite temperature and chemical potential in Sec.~\ref{sec:02_setup}. Then, we numerically solved these configurations, and the potential of $\rm{QQq}$ is discussed in Sec.~\ref{sec:03}. Finally, the main results of this paper are summarized in Sec.~\ref{sec:04}.

\section{The Setup}\label{sec:02_setup}
We start with a general metric ansatz\cite{Andreev:2015riv,Xun chen:2022bq}
\begin{equation}
\
d \mathbf{s}^{2}=\mathrm{e}^{s r^{2}} \frac{R^{2}}{r^{2}}\left(f(r)d t^{2}+d \vec{x}^{2}+ f^{-1}(r) d r^{2}\right)+\mathrm{e}^{-s r^{2}} g_{a b}^{(5)} d \omega^{a} d \omega^{b}
\end{equation}
This model parametrized by $s$ is a one-parameter deformation of Euclidean ${\rm AdS_5}$ space and a five-dimensional compact space (sphere) $\mathbf{X}$ whose coordinates are $\omega^a$. The radius $R$ is a constant, and $f(r)$ is a blacken factor of the black hole. The Hawking temperature of the black hole is defined as
\begin{equation}
T=\frac{1}{4 \pi}\left|\frac{d f}{d r}\right|_{r=r_h}=\frac{1}{\pi r_h}\left(1-\frac{1}{2} Q^2\right).
\end{equation}
Here, with $Q=qr_h{^2}$ and $0\leq Q\leq \sqrt{2}$, $r_h$ is the position of the black hole horizon. When $r$ is small, the behavior of the bulk gauge field is $A_{0}(r)=\mu-\eta r^{2}$ with $\eta=\kappa q$, where $\kappa$ is a dimensionless parameter and $q$ is the charge of the black hole. Together with the boundary conditions $A_{0}(r_h)=0 $ and $ A_{0}(0)=\mu$, we can acquire the relation between the baryon chemical potential $\mu$ and black charge $q$, which can be expressed as

\begin{equation}
\mu=q \kappa r_h^2=\kappa \frac{Q}{r_h}.
\end{equation}

Following Refs.\cite{Andreev:2020pqy,Andreev:2015riv,Xun chen:2022bq}, we fix the parameter $\kappa$ to 1. Furthermore, we have following three relationships

\begin{equation}\label{tmurh}
\begin{aligned}
f(r)=1-\left(\frac{1}{r_h^4}+\frac{\mu^2}{r_h^2}\right) r^4+\frac{\mu^2}{r_h^4} r^6,\\
T=\frac{1}{\pi r_h}\left(1-\frac{1}{2} \mu^2 r_h^2\right),\\
A_{0}(r)=\mu-\mu\frac{r^2}{r_h^2}.
\end{aligned}
\end{equation}
Noticeably, $A_{0}(r)$ is a quantity that is dependent on $\mu$ as well as the horizon $r_h$.

As is similarly discussed in Refs.\cite{Xun chen:2022bq,Andreev:2020xor}, the string structure contains three parts: three quarks, the baryon vertex V and the strings among them. There are three basic ingredients to construct the string configuration.

Firstly, the Nambu-Goto action of a string is expressed as the following expression
\begin{equation}
S_{NG}=\frac{1}{2 \pi \alpha^{\prime}} \int_0^1 d \sigma \int_0^T d \tau \sqrt{\gamma},
\end{equation}
where $\gamma$ is an induced metric on the string world sheet.

Secondly, the baryon vertex is considered to be a five-brane according to the AdS/CFT dictionary. Since we are interested in a static quark potential, we choose a static gauge $\xi^0 = t$ and $\xi^a = \theta^a$ with $\theta^a$ coordinates on $\textbf{X}$. The action of the baryon vertex is given by
\begin{equation}
S_{\mathrm{vert}}=\tau_v \int d t \frac{\mathrm{e}^{-2 s r^2}}{r} \sqrt{f(r)},
\end{equation}
where $\tau_v$ is a dimensionless parameter defined by $\tau_{v}=\mathcal{T}_{5} R \operatorname{vol}(\mathbf{X})$ and $  \operatorname{vol}(\mathbf{X})$ is a volume of $\textbf{X}$\cite{Xun chen:2022bq,Andreev:2015iaa,Andreev:2020pqy}.

Thirdly, the ingredient of the light quark at the string endpoint is a sigma-model action in a tachyon background. As illustrated in Ref.\cite{Xun chen:2022bq}, the action $S_q$ could be written as
\begin{equation}
S_{\mathrm{q}}=\mathrm{m} \int d t \frac{\mathrm{e}^{\frac{s}{2} r^2}}{r} \sqrt{f(r)},
\end{equation}
where $\mathrm{m} $ =$R T_0$. This is the action of a point particle of mass $T_0$ at rest. The model parameters need to be fixed as follows:
$\mathbf{g} =\frac{R^2}{2 \pi \alpha^{\prime}}$, $k =\frac{\tau_v}{3\mathbf{g}} $ and $ n =\frac{m}{\mathbf{g}}$.

Finally, when the string endpoints with attached quarks couple to a background gauge field\cite{Andreev:2015riv,Andreev:2019cbc}, the world sheet action includes boundary terms that could be given by
\begin{equation}
S_{\mathrm{A}}=\mp \frac{1}{3} \int d t \mathrm{~A}_0.
\end{equation}
The minus and plus signs correspond to a quark and an antiquark.

As separate distance $L$ grows, there are three configurations in the model.

\subsection{Small $L$}
\begin{figure}
	\centering
	\includegraphics[width=12cm]{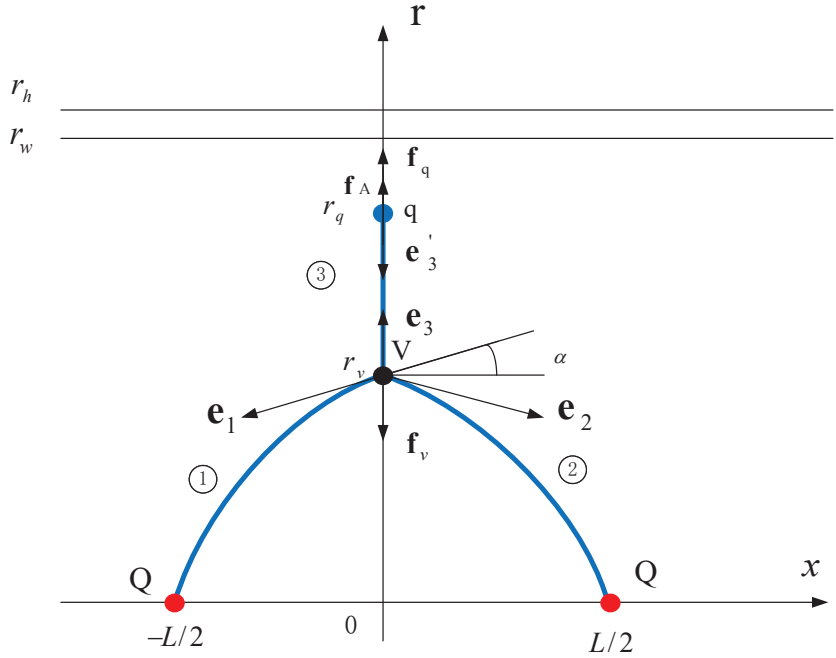}
	\caption{\label{small}The first static string configuration at a small heavy quark separation. The heavy quarks Q are placed at the ${x}$-axis. The light quark and baryon vertex V are on the $r$-axis at $r=r_q$ and $r=r_v$, respectively. The quarks and baryon vertex are connected by blue strings. The force exerted on the vertex and light quark are depicted by the black arrows. $r_h$ is the position of the black-hole horizon. $r_w$ is the position of the dynamic wall.}
\end{figure}

The first configuration is plotted in Fig.~\ref{small}. The total action is the sum of the Nambu-Goto actions, the action for the vertex, light quark and boundary terms
\begin{equation}
S=\sum_{i=1}^3 S_{\mathrm{NG}}^{(i)}+S_{\mathrm{vert}}+S_{\mathrm{q}}+\left.2S_{\mathrm{A}}\right|_{r=r_Q}+\left.S_{\mathrm{A}}\right|_{r=r_q}.
\end{equation}
If we choose the static gauge $\xi^{1}=t$ and $\xi^{2}=r$, then the boundary conditions for $x(r)$ are
\begin{equation}
x^{(1)}(0)=-\frac{L}{2} , \quad x^{(2)}(0)=\frac{L}{2}, \quad x^{(i)}\left(r_{v}\right)=x^{(3)}\left(r_{q}\right)=0.
\end{equation}
The total action can be written as
\begin{equation}\label{totalaction1}
\begin{aligned}
S&=\mathbf{g} T \Big(2 \int_{0}^{r_{v}} \frac{d r}{r^{2}}  \mathrm{e}^{s r^{2}} \sqrt{1+f(r)\left(\partial_{r} x\right)^{2}}\\&+\int_{r_{v}}^{r_{q}} \frac{d r}{r^{2}} \mathrm{e}^{\mathrm{s} r^{2}} \sqrt{1+f(r)\left(\partial_{r} x\right)^{2}}+3 \mathrm{k} \frac{\mathrm{e}^{-2 \mathrm{~s} r_{v}^{2}}}{r_{v}}\sqrt{f(r_v)}\\&+\mathrm{n} \frac{\mathrm{e}^{\frac{1}{2} \mathrm{~s} r_{q}^{2}}}{r_{q}}\sqrt{f(r_q)}\Big)-\frac{1}{3}A_0(r_q)-\frac{2}{3}A_0(0),
\end{aligned}
\end{equation}
where $\partial_{r} x=\frac{\partial x}{\partial r}$ and $T=\int_{0}^{T} d t$. We consider the first term in Eq.~(\ref{totalaction1}) which corresponds to string \textcircled{1} and string \textcircled{2}. Due to symmetry, the result of the string \textcircled{2} is exactly the same as that of the string \textcircled{1}. The equation of motion for $x(r)$ can be obtained from the Euler-Lagrange equation. It reduces to
\begin{equation}
\mathcal{I}=\frac{w(r) f(r) \partial_{r} x}{\sqrt{1+f(r)\left(\partial_{r} x\right)^{2}}}, \quad w(r)=\frac{\mathrm{e}^{\mathrm{s} r^{2}}}{r^{2}}.
\end{equation}
$\mathcal{I}$ is a constant. At $r_v$, we have $\left.\partial_{r} x\right|_{r=r_{v}}= \cot \alpha$ with $\alpha>0$ and
\begin{equation}
\mathcal{I}=\frac{w(r_v) f(r_v) \cot \alpha}{\sqrt{1+f(r)\cot^{2} \alpha}}.
\end{equation}
$\partial_{r} x$ can be solved as
\begin{equation}\label{xp1}
\partial_{r} x = \sqrt{\frac{\omega\left(r_{v}\right)^{2} f\left(r_{v}\right)^{2}}{\left(f\left(r_{\nu}\right)+ \tan ^{2} \alpha\right) \omega(r)^{2} f(r)^{2}-f(r) w\left(r_{v}\right)^{2} f(r_v)^{2}}}.
\end{equation}
By virtue of Eq.~(\ref{xp1}), the integral over [$0,r_v$] of $dr$ is

\begin{equation}\label{distance1}
L=2 \int_{0}^{r_{v}} \frac{dx}{dr} dr.
\end{equation}
The regularized energy after subtracting the divergent term $\mathbf{g} \int_{0}^{\infty}dr\frac{1}{r^2}$ from the first term of Eq.~(\ref{totalaction1}) can be expressed as
\begin{equation}\label{energy1}
E_{1}=\frac{S}{T}=\mathbf{g} \int_{0}^{r_{v}}(\frac{1}{r^{2}}  \mathrm{e}^{\mathbf{s} r^{2}} \sqrt{1+f(r)\left(\partial_{r} x\right)^{2}}-\frac{1}{r^2})d r - \frac{\mathbf{g}}{r_v} + c.
\end{equation}
Here $c$ is a normalization constant. As string \textcircled{3} is straight stretched between the vertex and light quark from Eq.~(\ref{totalaction1}), the energy of which could be calculated as
\begin{equation}
E_{2}=\mathbf{g} \int_{r_{v}}^{r_{q}}\frac{ \mathrm{e}^{\mathbf{s} r^{2}}}{r^{2}}  d r.
\end{equation}
To this end, the energy could be written as

\begin{equation}\label{freeenergy1}
\begin{aligned}
E_{\mathrm{QQq}}&=\mathbf{g} \Big(2\int_{0}^{r_{v}}(\frac{1}{r^{2}}  \mathrm{e}^{\mathbf{s} r^{2}} \sqrt{1+f(r)\left(\partial_{r} x\right)^{2}}-\frac{1}{r^2})d r - \frac{2}{r_v}\\& + \mathrm{n} \frac{\mathrm{e}^{\frac{1}{2} s r_q^2}}{r_q} +3 \mathrm{k} \frac{\mathrm{e}^{-2 s r_v^2}}{r_v}\sqrt{f(r_v)}\Big)-\frac{1}{3}A_0(r_q)-\frac{2}{3}A_0(0)+2 c.
\end{aligned}
\end{equation}

It is easy to find that the potential energy is a function of $r_v$,$r_h$ and angle $\alpha$. As discussed before, $\mu$, $T$ and $r_h$ have a relationship as Eq.~(\ref{tmurh}). Therefore, one can study the potential of $\rm{QQq}$ in a model associated with the temperature and chemical potential. The main point becomes to determine $r_q$(the position of the light quark) and angle $\alpha$. Using the force balance equation with respect to $r_q$ and $r_v$, we can achieve this goal.

Firstly, the force balance equation at $r_q$ is
\begin{equation}
f_{\mathrm{q}}+e_{3}^{\prime}+f_A=0.
\end{equation}
By varying the action for $r_q$, it is found that $f_{\mathrm{q}}=\Big(0,-\mathbf{g} n \partial_{r_q}(\frac{\mathrm{e}^{\frac{1}{2} s r_q^2}}{r_q}\sqrt{f(r_q)})\Big)$, $f_A=\frac{\partial {A_0}(r)}{\partial r}$ and the string tension  $\mathbf{e}_{3}^{\prime}=\mathbf{g} w\left(r_{q}\right)(0,-1)$.
Therefore, the balance equation becomes
\begin{equation}\label{balance equation 1}
\begin{aligned}
&2 {r_q}^2 \sqrt{f({r_q})} {A_0}'({r_q})-3 \mathbf{g} n e^{\frac{{r_q}^2 s}{2}} \left({r_q} f'({r_q})+2 f({r_q}) \left({r_q}^2 s-1\right)\right)-6 \mathbf{g} \sqrt{f({r_q})} e^{{r_q}^2 s}=0.
\end{aligned}
\end{equation}
With a suitable $T$ and $\mu$ fixed, $r_h$ can be solved through  Eq.~(\ref{tmurh}). By solving Eq.~(\ref{balance equation 1}), we find that $r_q$ only depends on $r_h$. Hereafter, turning to the angle $\alpha$, the force balance equation on the vertex can be expressed as
\begin{equation}\label{force balance}
\mathbf{e}_{1}+\mathbf{e}_{2}+\mathbf{e}_{3}+\mathbf{f}_{v}=0.
\end{equation}

We adopt the same manner as \cite{Andreev:2021bfg}, and we get such a result, the force on the vertex is  $\mathbf{f}_{v}=\left(0,-3 \mathbf{g} k \partial_{r_{v}} (\frac{\mathrm{e}^{-2 s r_{v}^{2}}}{r_{v}}\sqrt{f(r_v)})\right)$ and string tensions are $\mathbf{e}_{1}=\mathbf{g} w\left(r_{v}\right)(-\frac{f(r_{v})}{\sqrt{\tan^2 \alpha + f(r_{v})}},-\frac{1}{\sqrt{f(r_{v}) \cot^2 \alpha + 1}})$, $\mathbf{e}_{2}=\mathbf{g} w\left(r_{v}\right)(\frac{f(r_{v})}{\sqrt{\tan^2 \alpha + f(r_{v})}},-\frac{1}{\sqrt{f(r_{v}) \cot^2 \alpha + 1}})$, $\mathbf{e}_{3}=\mathbf{g} w\left(r_{v}\right)(0,1)$. The nontrivial component of the force balance equation is
\begin{equation}\label{force balance equation 1}
\frac{3 \mathbf{g} k e^{-2 {r_v}^2 s} \left(f({r_v}) \left(8 {r_v}^2 s+2\right)-{r_v} f'({r_v})\right)}{\sqrt{f({r_v})}}+2 \mathbf{g} e^{{r_v}^2 s} \Big(1-\frac{2}{\sqrt{\cot ^2{\alpha } f({r_v})+1}}\Big)=0.
\end{equation}
The Eq.~(\ref{force balance equation 1}) provides us with a relationship between $r_v$  and $\alpha$. Then, we can numerically solve the energy $E$ and separation distance $L$.

\subsection{Intermediate $L$}
\begin{figure}
	\centering
	\includegraphics[width=12cm]{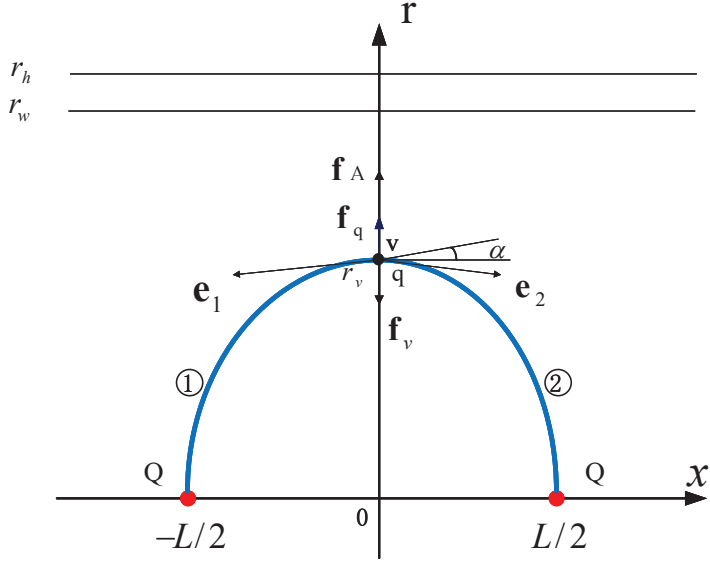}
	\caption{\label{intermediate}The second static string configuration at an intermediate heavy quark separation. The heavy quarks Q are placed at the $x$-axis. The light quark and baryon vertex V are on the $r$-axis at the same point. The quarks and baryon vertex are connected by blue strings. The force exerted on the vertex and light quark are depicted by the black arrows. $r_h$ is the position of the black-hole horizon. $r_w$ is the position of the dynamic wall.}
\end{figure}

In this configuration, the light quark coincides with the baryon vertex, and there are no strings between them. The angle $\alpha$ will gradually decrease to zero as the separation distance $L$ increases. Thus, the total action of the configuration plotted in Fig.\ref{intermediate} can be expressed as
\begin{equation}
S=\sum_{i=1}^2 S_{\mathrm{NG}}^{(i)}+S_{\mathrm{vert}}+S_{\mathrm{q}}+\left.2S_{\mathrm{A}}\right|_{r=r_Q}+\left.S_{\mathrm{A}}\right|_{r=r_q}.
\end{equation}

The static gauge is set in the same way as before. Thus, the total energy of the configuration is

\begin{equation}\label{freeenergy2}
\begin{aligned}
E_{\mathrm{QQq}}&=\mathbf{g} \Big(2\int_{0}^{r_{v}}(\frac{1}{r^{2}}  \mathrm{e}^{\mathbf{s} r^{2}} \sqrt{1+f(r)\left(\partial_{r} x\right)^{2}}-\frac{1}{r^2})d r - \frac{2}{r_v}+ \mathrm{n} \frac{\mathrm{e}^{\frac{1}{2} s r_v^2}}{r_v} +3 \mathrm{k} \frac{\mathrm{e}^{-2 s r_v^2}}{r_v}\sqrt{f(r_v)}\Big)\\& -\frac{1}{3}A_0(r_q)-\frac{2}{3}A_0(0)+2 c.
\end{aligned}
\end{equation}
The only difference between the above equation with Eq.~(\ref{freeenergy1}) is that $r_q$ is replaced by $r_v$. Similarly, at $r_q$ (or $r_v$), we have
\begin{equation}\label{forcebalance2}
\mathbf{e}_{1}+\mathbf{e}_{2}+\mathbf{f}_{v}+\mathbf{f}_{q}+\mathbf{f}_{A}=0.
\end{equation}
Each force could be given by
$$\mathbf{f}_{q}=\left(0,-n \mathbf{g}  \partial_{r_{v}}\left(\frac{\mathrm{e}^{\frac{1}{2} s r_{v}^{2}}}{r_{v}} \sqrt{f\left(r_{v}\right)}\right)\right),$$

$$\mathbf{f}_{v}=\left(0,-3 \mathbf{g}  k \partial_{r_{v}}\left(\frac{\mathrm{e}^{-2 s r_{v}^{2}}}{r_{v}} \sqrt{f\left(r_{v}\right)}\right)\right),$$

$$\mathbf{e}_{1}=\mathbf{g} w\left(r_{v}\right)(-\frac{f(r_{v})}{\sqrt{\tan^2 \alpha + f(r_{v})}},-\frac{1}{\sqrt{f(r_{v}) \cot^2 \alpha + 1}}),$$

$$\mathbf{e}_{2}=\mathbf{g} w\left(r_{v}\right)(\frac{f(r_{v})}{\sqrt{\tan^2 \alpha + f(r_{v})}},-\frac{1}{\sqrt{f(r_{v}) \cot^2 \alpha + 1}}),$$

$$\mathbf{f}_{A}=\frac{\partial {A_0}(r)}{\partial r} .$$

The force balance equation can be rearranged as
\begin{equation}
\begin{aligned}
&\frac{2 {r_v}^2 e^{2 {r_v}^2 s} {A_0}'({r_v})}{3 \mathbf{g}}+\frac{3 k \left(f({r_v}) \left(8 {r_v}^2 s+2\right)-{r_v} f'({r_v})\right)}{\sqrt{f({r_v})}}\\&-\frac{n e^{\frac{5 {r_v}^2 s}{2}} \left({r_v} f'({r_v})+2 f({r_v}) \left({r_v}^2 s-1\right)\right)}{\sqrt{f({r_v})}}-\frac{4 e^{3 {r_v}^2 s}}{\sqrt{\cot {\alpha }^2 f({r_v})+1}}=0.
\end{aligned}
\end{equation}

\subsection{Large $L$}
\begin{figure}
	\centering
	\includegraphics[width=12cm]{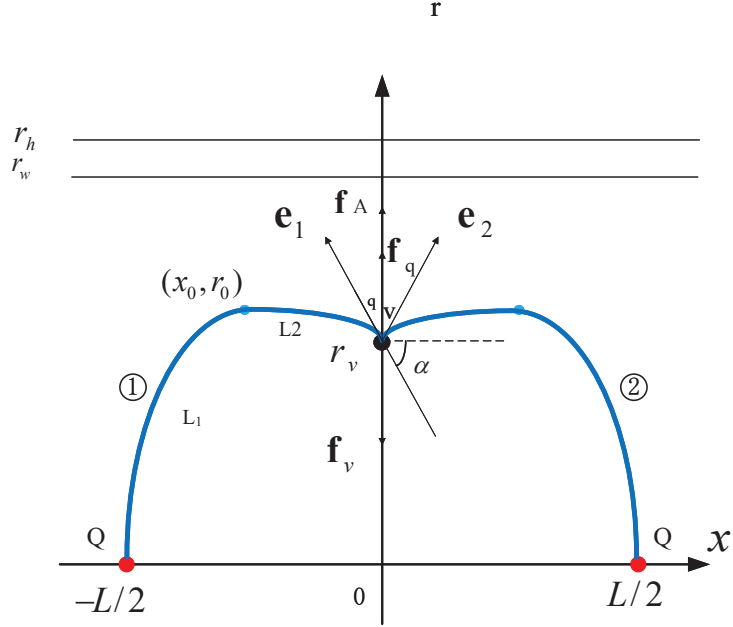}
	\caption{\label{large}The third static string configuration at a large heavy quark separation as the “U-shape” string transforms to an “M-shape” string. The heavy quarks Q are placed at the $x$-axis. The light quark and baryon vertex V are on the $r$-axis at $r=r_q=r_v$. The quarks and baryon vertex are connected by blue strings. The force exerted on the vertex and light quark are depicted by the black arrows. $r_h$ is the position of the black-hole horizon. $r_w$ is the position of the dynamic wall.}
\end{figure}

In this section, with the increase of $r_v$, the U-shape string develops to an M-shape as in Fig.\ref{large}. To simplify the calculation, we choose another static gauge  $\xi^{1}=t$ and $\xi^{2}=x$; the boundary conditions are
\begin{equation}
r^{(1)}(-L / 2)=r^{(2)}(L / 2)=0, \quad r^{(i)}(0)=r_{v} .
\end{equation}
The total action becomes
\begin{equation}\label{Stotal3}
\begin{aligned}
S&=\mathbf{g} T\Big(\int_{-L / 2}^{0} d x w(r) \sqrt{f(r)+\left(\partial_{x} r\right)^{2}}+\int_{0}^{L / 2} d x w(r) \sqrt{f(r)+\left(\partial_{x} r\right)^{2}}+3 \mathrm{k} \frac{\mathrm{e}^{-2 s r_{v}^{2}}}{r_{v}}\sqrt{f(r_v)}\\
&+\mathrm{n} \frac{\mathrm{e}^{\frac{1}{2} \mathrm{~s} r_{v}^{2}}}{r_{v}}\sqrt{f(r_v)}\Big)-\frac{1}{3}A_0(r_v)-\frac{2}{3}A_0(0).
\end{aligned}
\end{equation}

The action of the string could be given by the first two terms of Eq.~(\ref{Stotal3}). From the force balance equation at the baryon vertex, angle $\alpha$ can be similarly calculated in this configuration. In addition, we  can use the first integral at points $r_{0}$ and $r_{v}$ to obtain the relationship among $r_0$, $r_v$, and $\alpha$,
\begin{equation}\label{first integral}
\begin{aligned}
\frac{w(r)f(r)}{\sqrt{f(r)+\left(\partial_{x} r\right)^{2}}}=w(r_0)\sqrt{f(r_0)},\\
\frac{w(r_v)f(r_v)}{\sqrt{f(r_v)+ \tan\alpha^2}}=w(r_0)\sqrt{f(r_0)}.
\end{aligned}
\end{equation}
After a short calculation, the solution for $\partial_x r$  can be written as
\begin{equation}\label{first}
\partial_x r =\sqrt{\frac{w(r)^2f(r)^2f(r_0) - f(r)w(r_0)^2 f(r_0)^2}{w(r_0)^2 f(r_0)^2}}.
\end{equation}
The separate distance consists of two parts separated by the turning point $(x_0,r_0)$,
\begin{equation}\label{distance2}
\begin{aligned}
L &= 2(L_1 + L_2) = 2\Big(\int_0^{r_0} \frac{dx}{dr} dr + \int_{r_v}^{r_0} \frac{dx}{dr} dr\Big)\\
&=2\Big(\int_0^{r_0} \sqrt{\frac{w(r_0)^2 f(r_0)^2}{w(r)^2f(r)^2f(r_0) - f(r)w(r_0)^2f(r_0)^2}} dr \\
&+ \int_{r_v}^{r_0} \sqrt{\frac{w(r_0)^2 f(r_0)^2}{w(r)^2f(r)^2f(r_0) - f(r)w(r_0)^2f(r_0)^2}} dr\Big).
\end{aligned}
\end{equation}
Subtracted the divergent terms, the renormalized energy of string \textcircled{1} is
\begin{equation}
\begin{aligned}
E &=\mathbf{g} \int_{r_v}^{r_0}  \Big(w(r) \sqrt{\frac{w(r)^2 f(r)^2 f(r_0)}{w(r)^2f(r)^2f(r_0) - f(r) w(r_0)^2 f(r_0)^2}}\Big) dr \\&+\mathbf{g} \int_0^{r_0} \Big( w(r) \sqrt{\frac{w(r)^2 f(r)^2 f(r_0)}{w(r)^2f(r)^2f(r_0) - f(r) w(r_0)^2 f(r_0)^2}}- \frac{1}{r^2}\Big) dr - \frac{\mathbf{g}}{r_0}+ 2c.
\end{aligned}
\end{equation}
Since string \textcircled{1} and string \textcircled{2} are completely symmetric, the calculation method is the same, whereupon the total energy of the final configuration can be written as
\begin{equation}
\begin{aligned}
E_{\mathrm{QQq}} &=\mathbf{g} \Big(  2\int_{r_v}^{r_0} w(r) \sqrt{\frac{w(r)^2 f(r)^2 f(r_0)}{w(r)^2f(r)^2f(r_0) - f(r) w(r_0)^2 f(r_0)^2}}dr+\\ & 2\int_0^{r_0} (w(r) \sqrt{\frac{w(r)^2 f(r)^2 f(r_0)}{w(r)^2f(r)^2f(r_0) - f(r) w(r_0)^2 f(r_0)^2}} - \frac{1}{r^2}) dr
\\& - \frac{1}{r_0}+ 3 \mathrm{k} \frac{\mathrm{e}^{-2 \mathrm{~s} r_{v}^{2}}}{r_{v}}\sqrt{f({r_v})}+\mathrm{n} \frac{\mathrm{e}^{\frac{1}{2} \mathbf{s} r_{v}^{2}}}{r_{v}}\sqrt{f({r_v})}\Big) + 2c.
\end{aligned}
\end{equation}
Now that all the variables in the integral have been represented by $r_v$, we can numerically calculate the results of $L$ and $E_{\mathrm{QQq}}$.

\section{NUMERICAL RESULT AND DISCUSSION}\label{sec:03}

The string-breaking behavior can be judged from the total potential energy. Besides, the system will melt when $T$ and $\mu$ become large enough. During the process of string melting, we compare the potential energies of doubly flavor baryons to those of quark-antiquark pairs at fixed temperature and chemical potential, respectively. String melting can be judged  from the $L-r_{v}$ diagram. Within the margin of numerical error, we draw a $T-\mu$ melting diagram.

To approach the lattice result at vanishing temperature and chemical potential, all the parameters are fixed at vanishing temperature and chemical potential as follows: $s$ = 0.42 $\rm{GeV^2}$, $\mathbf{g} = 0.176$, $n = 3.057$ and $c$ = 0.623 GeV\cite{Andreev:2020xor}.

\subsection{The numerical analysis of three configurations}
\begin{figure}
	\centering
	\includegraphics[width=8cm]{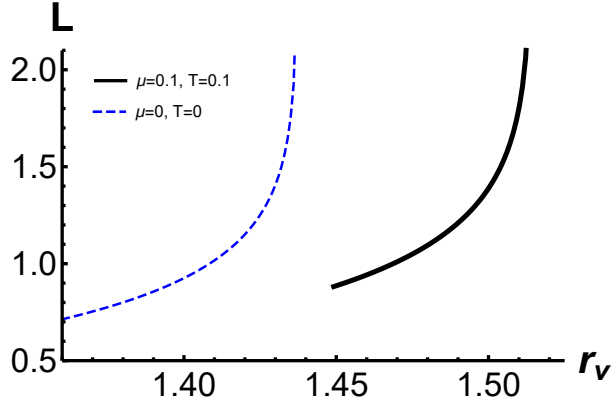}
	\caption{\label{Lrv0001} At large separate distances, $L$ as a function of $r_v$. The solid black and dashed blue lines correspond to the variations of $L$ as a function of $r_v$ for $\rm{QQq}$ at $\mu$=0.1 GeV, $T$=0.1 GeV, and $\mu$=0 GeV, $T$=0 GeV, respectively. The unit of $L$ is fm,  $r_v$ is $\rm{GeV^{-1}}$.}
\end{figure}

\begin{figure}
	\centering
	\includegraphics[width=8cm]{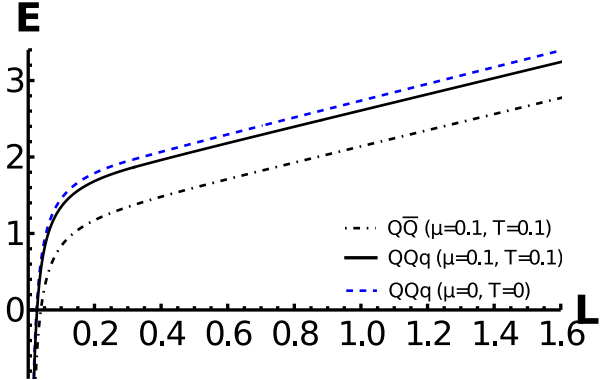}
	\caption{\label{EL} $E$ as a function of $L$ for $\rm{QQq}$ and $\rm{Q\bar{Q}}$. The solid black and dashed blue lines correspond to the variations of $E$ as a function of $L$ for $\rm{QQq}$ at $\mu$=0.1 GeV, $T$=0.1 GeV, and $\mu$=0 GeV, $T$=0 GeV, respectively. The dash-dotted black line represents the variations for $\rm{Q\bar{Q}}$ at $\mu$=0.1 GeV, $T$=0.1 GeV. The unit of  $E$ is GeV, and $L$ is fm.}
\end{figure}

According to the previous calculation procedure, the  diagram of $L-r_v$ ($\mu$=0.1 GeV, $T$=0.1 GeV) and $E-L$  before string melting can be shown in Fig.~\ref{Lrv0001} and Fig~\ref{EL}. In Fig.~\ref{Lrv0001}, it is shown that the separate distance will increase with the increase of $r_v$. The presence of temperature and chemical potential will cause that the separate distance will tend to infinite at large $r_v$. In Fig.~\ref{EL}, we plot the potential energy of the quark-antiquark pair with a dot-dashed line and compare with that of doubly heavy baryon under the same external conditions. While the overall behavior of potential energy also follows a Cornell-like potential, $\rm{Q \bar{Q}}$ has a lower potential energy at equal separate distances. Besides, it is shown that the presence of temperature and chemical potential will decrease the linear part of the Cornell-like potential.

In addition, despite $\mu$ having been considered in the model, the potential energy is still the Coulomb potential in small $L$(the first configuration) and linear potential in intermediate and large $L$(the later two configurations). Theoretically, $L$ can grow to infinity as $r_v$ increases if no string melting occurs.

\begin{figure}
	\centering
	\includegraphics[width=8cm]{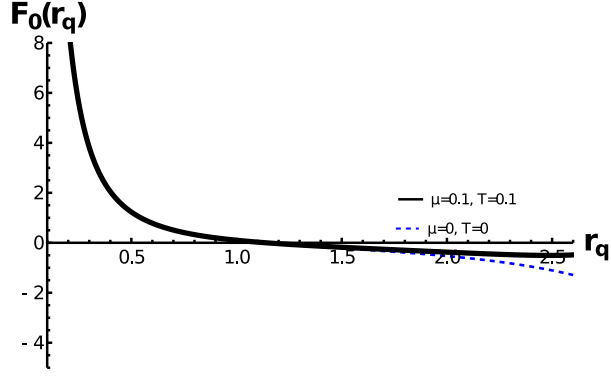}
	\caption{\label{configurationexist} The force balance equations as functions of $r_q$. The solid black and dashed blue lines correspond to $\mu$=0.1 GeV, $T$=0.1 GeV, and $\mu$=0 GeV, $T$=0 GeV, respectively. $r_q$ is the position of light quark. The unit of $r_q$ is $\rm{GeV^{-1}}$.}
\end{figure}

\begin{figure}
	\centering
	\includegraphics[width=8cm]{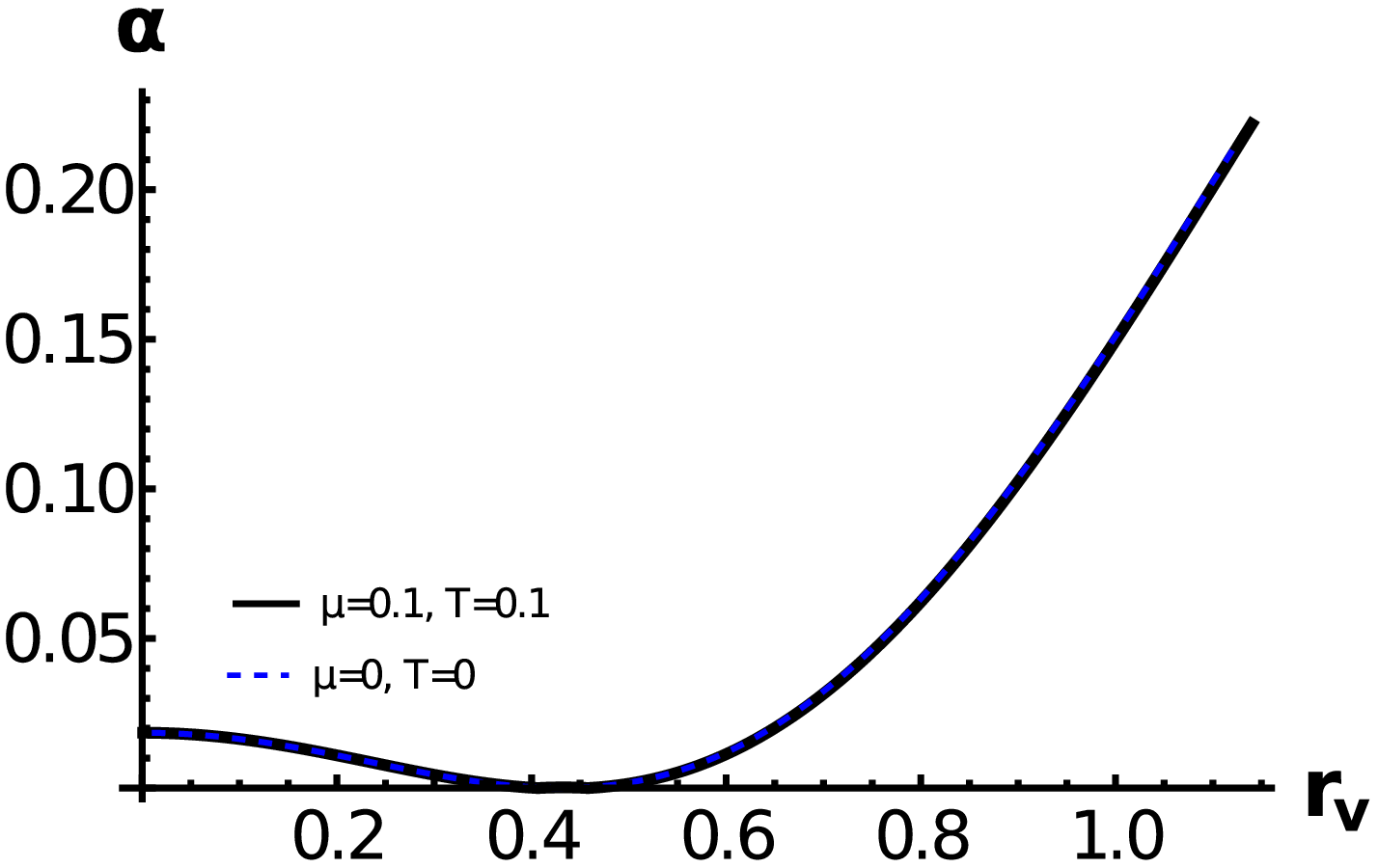}
	\caption{\label{figure1} $\alpha$ as a function of $r_v$ for $\rm{QQq}$ in small separate distance. The solid black and dashed blue lines correspond to $\mu$=0.1 GeV, $T$=0.1 GeV, and $\mu$=0 GeV, $T$=0 GeV, respectively. The unit of $r_v$ is $\rm{GeV^{-1}}$.}
\end{figure}

\begin{figure}
	\centering
	\includegraphics[width=8cm]{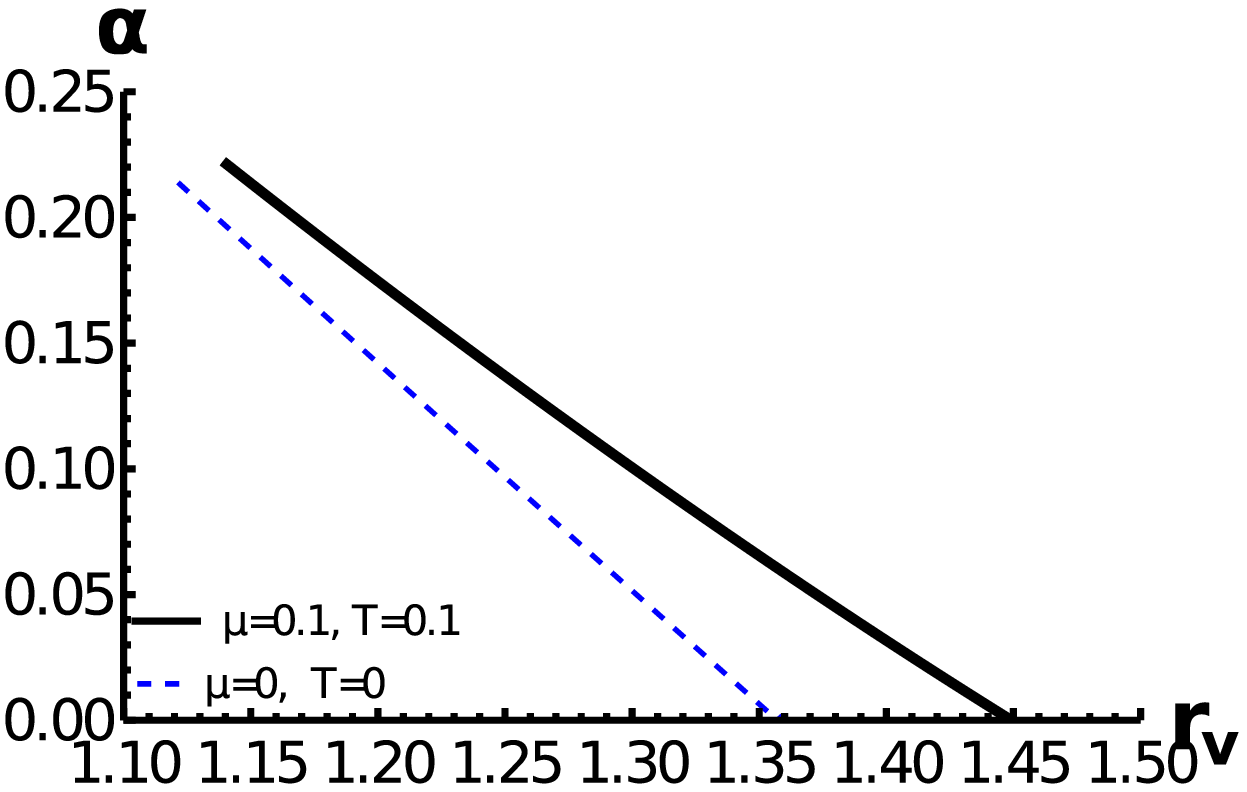}
	\caption{\label{figure2} $\alpha$ as a function of $r_v$ for $\rm{QQq}$ in intermediate separate distance.The solid black and dashed blue lines correspond to $\mu$=0.1 GeV, $T$=0.1 GeV, and $\mu$=0 GeV, $T$=0 GeV, respectively. The unit of $r_v$ is $\rm{GeV^{-1}}$.}
\end{figure}

\begin{figure}
	\centering
	\includegraphics[width=8cm]{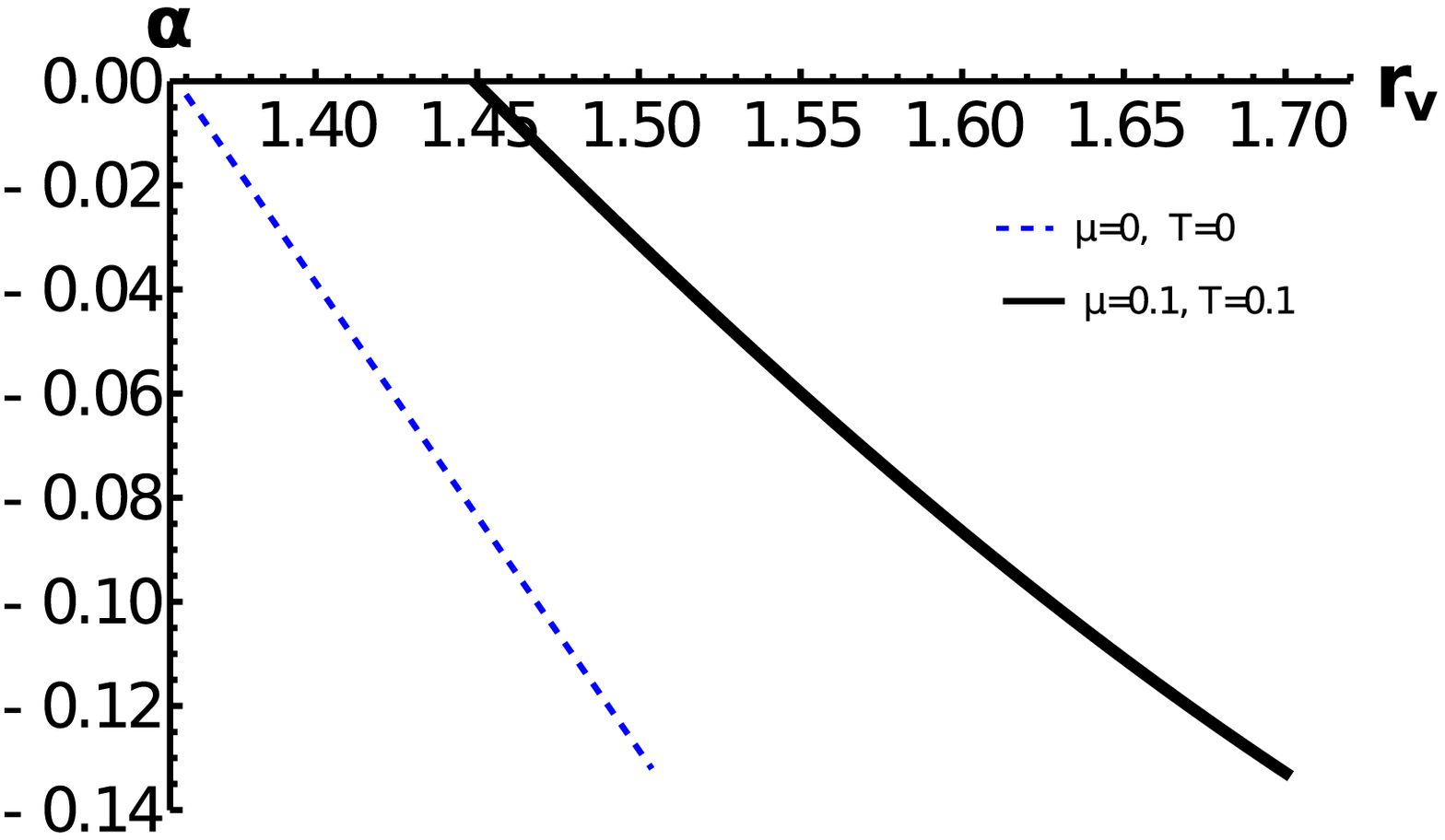}
	\caption{\label{figure3} $\alpha$ as a function of $r_v$ for $\rm{QQq}$ in large separate distance. The solid black and dashed blue lines correspond to $\mu$=0.1 GeV, $T$=0.1 GeV, and $\mu$=0 GeV, $T$=0 GeV, respectively. The unit of $r_v$ is $\rm{GeV^{-1}}$.}
\end{figure}

When $T$=0.1 GeV, $\mu$=0.1 GeV, all string configurations exist. According to Eq.~(\ref{balance equation 1}), the force balance equation of the light quark  $F_0(r_q)$ is plotted  in Fig.~\ref{configurationexist}. At the same time, we further gain understanding of the change for string structure by observing the variation for angle $\alpha$. For the three configurations mentioned above, we solve the force balance equations at the baryon vertex to obtain a plot of the variation of angle $\alpha$ with $r_v$ for the three configurations as shown in Fig.~\ref{figure1}, Fig.~\ref{figure2} and Fig.~\ref{figure3}, respectively. At small separate distance, $\alpha$ first decreases and then increases as $r_v$ increases. At intermediate separate distance, $\alpha$ slowly decreases from $\alpha\approx 0.221$ to zero as $r_{v}$ increases. At large separate distance, as $r_{v}$ increases, angle $\alpha$ becomes negative and continues to decrease monotonically. Besides, it is shown that the temperature and chemical potential have minimal effect at small separate distances.

\subsection{String breaking}
\begin{figure}
	\centering
	\includegraphics[width=8cm]{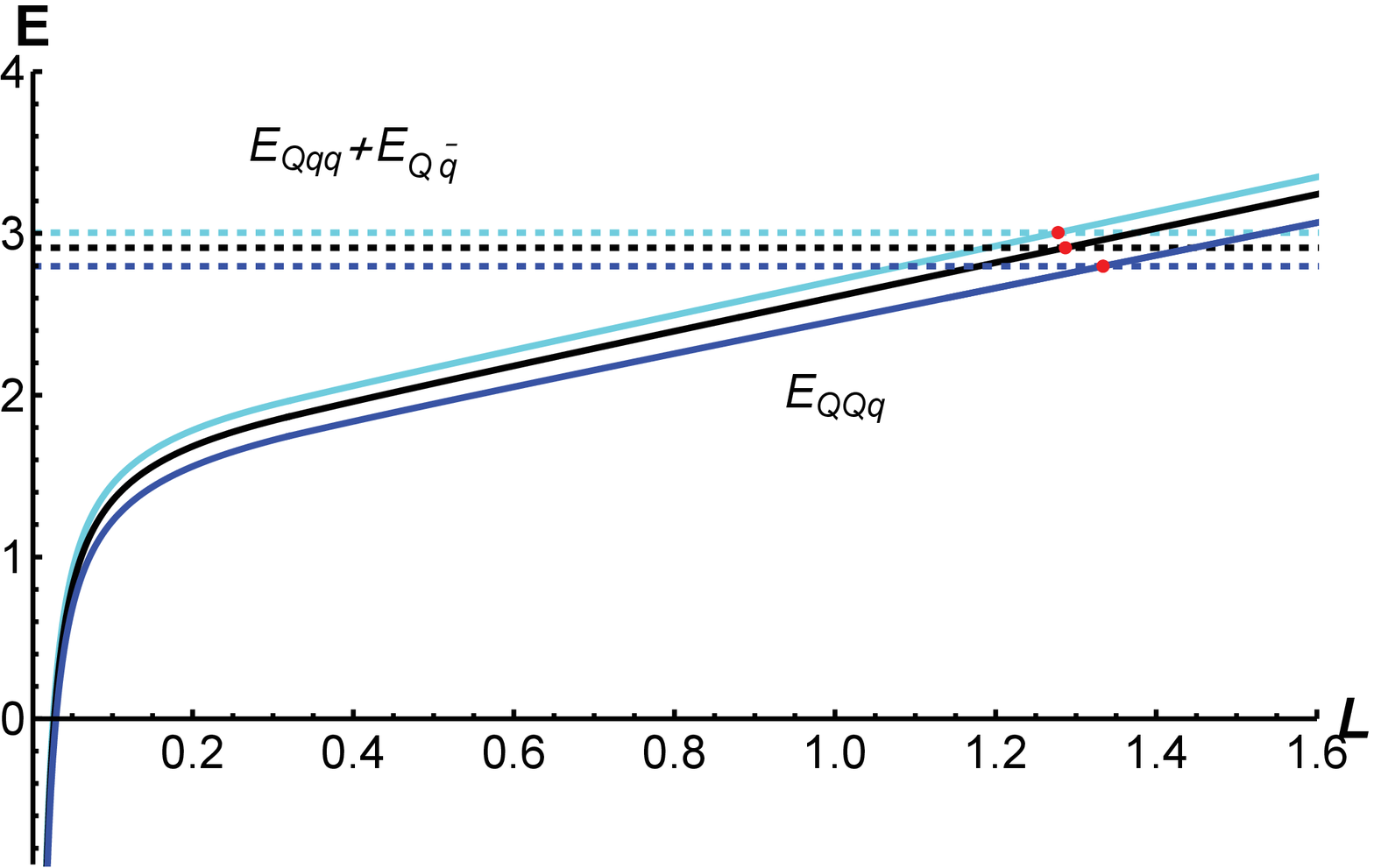}
	\caption{\label{suilie} $E$ as a function of $L$. The cyan line, the black line, and the blue line  represent $\mu$=0 GeV, $\mu$=0.1 GeV and $\mu$=0.22 GeV, respectively. $T$ remains 0.1 GeV.  The dashed line shows the potential energy after decay. The unit of $E$ is GeV, and $L$ is fm.}
\end{figure}

\begin{figure}
	\centering
	\includegraphics[width=8cm]{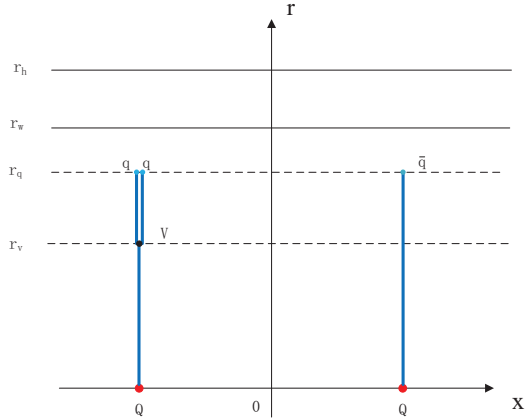}
	\caption{\label{decay} A schematic diagram of $\rm{Q q q+Q \bar{q}}$.}
\end{figure}

In this subsection, we  fixed $T$=0.1 GeV, and set $\mu$ = 0 GeV, 0.1 GeV, and 0.22 GeV, respectively.
As is well known, due to the restriction of energy, string breaking will occur at a critical length  when the separation distance $L$ increases. At this point, the energy of the string reaches a critical level where it can no longer remain stable. It breaks apart and becomes two different string structures. In our paper, we are interested in the decay mode of
$\rm{Q Q q \rightarrow Q q q+Q \bar{q}}.$
The renormalized total energy is
\begin{equation}
\begin{aligned}
E_{\mathrm{QQq}}+E_{Q\bar{q}} &= \mathbf{g} \Big(2 \int_{r_v}^{r_q} \frac{e^{s r^2}}{r^2} dr + \int_{0}^{r_q}(\frac{{e^{s r^2}}}{r^2} - \frac{1}{r^2}) - \frac{1}{r_q} +  \int_{0}^{r_v} (\frac{e^{s r^2}}{r^2}-\frac{1}{r^2})- \frac{1}{r_v}+ 3k  \frac{e^{-2 s r_v^2} \sqrt{f(r_v)}}{r_v}\\&+ 3 n  \frac{e^{\frac{1}{2}s r_q^2}}{r_q} \sqrt{f(r_q)}\Big) + 2c.
\end{aligned}
\end{equation}
Expressions for the breaking distances give characteristic scales when string breaking takes place\cite{Andreev:2019cbc,Andreev:2020pqy}. The potential energy and the schematic diagram of string breaking  are presented in Fig.~\ref{suilie} and Fig.~\ref{decay} respectively. As can be seen from Fig.~\ref{suilie}, with the increase of $\mu$, the energy of the breaking point decreases, and the separate distance increases. It means the $\rm{QQq}$ configuration is more stable at small chemical potential.

\subsection{String melting}
\begin{figure}
	\centering
	\includegraphics[width=8cm]{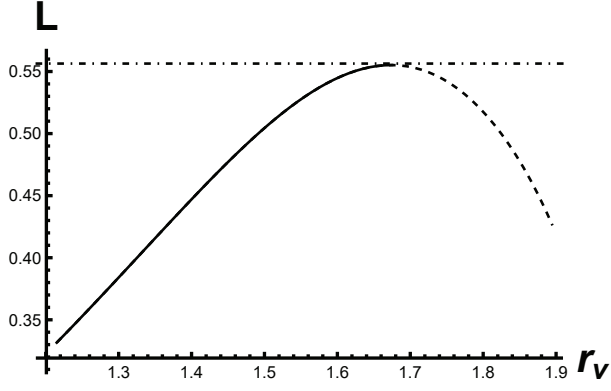}
	\caption{\label{beyondrw} $L$ as a function of $r_v$ at $\mu$=0.35 GeV and $T$=0.1 GeV. The part indicated by the dashed line means that the string has melted and has no physical significance. The dotted lines represent the maximum value of potential energy. The unit of $L$ is fm, and $r_v$ is $\rm{GeV^{-1}}$.}
\end{figure}

\begin{figure}
	\centering
	\includegraphics[width=8cm]{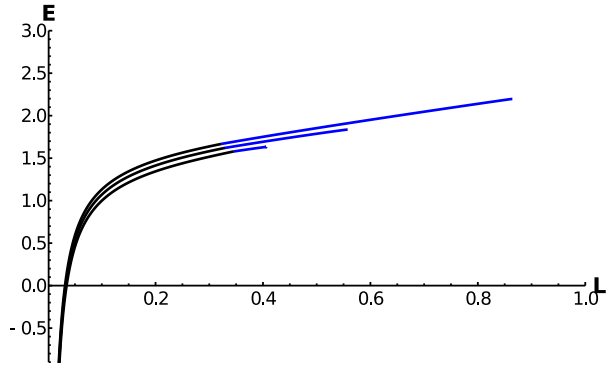}
	\caption{\label{melting}$E$ as a function of $L$ when $T$ is fixed to 0.1 GeV. From top to bottom, $\mu$ is 0.3 GeV, 0.35 GeV and 0.4 GeV, respectively. The unit of $E$ is GeV, and $L$ is fm.}
\end{figure}

\begin{figure}
	\centering
	\includegraphics[width=8cm]{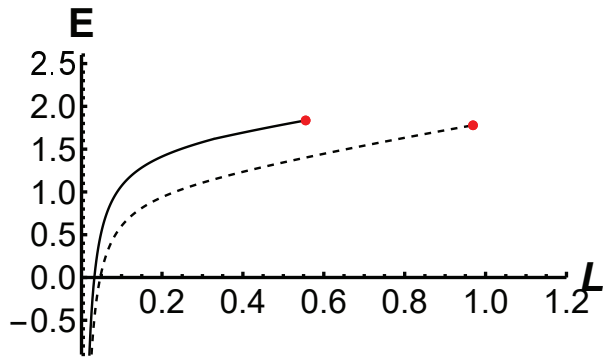}
	\caption{\label{t01} $E$ as a function of $L$ at $T$=0.1 GeV and $\mu$=0.35 GeV. The solid line and the dashed line correspond to $\rm{QQq}$ and $\rm{Q \bar{Q}}$ respectively. The unit of $E$ is GeV, and $L$ is fm.}
\end{figure}

\begin{figure}
	\centering
	\includegraphics[width=8cm]{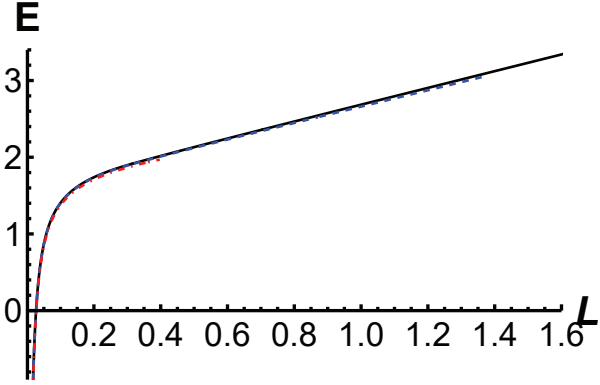}
	\caption{\label{mufixed}  $E$ as a function of $L$ at different $T$ when $\mu$=0.05 GeV. The black solid line, the blue dashed line, and the red dot-dashed line correspond to $T$=0.05 GeV, $T$=0.1 GeV, and $T$=0.146 GeV respectively. The unit of $E$ is GeV ,and $L$ is fm.}
\end{figure}

\begin{figure}
	\centering
	\includegraphics[width=16cm]{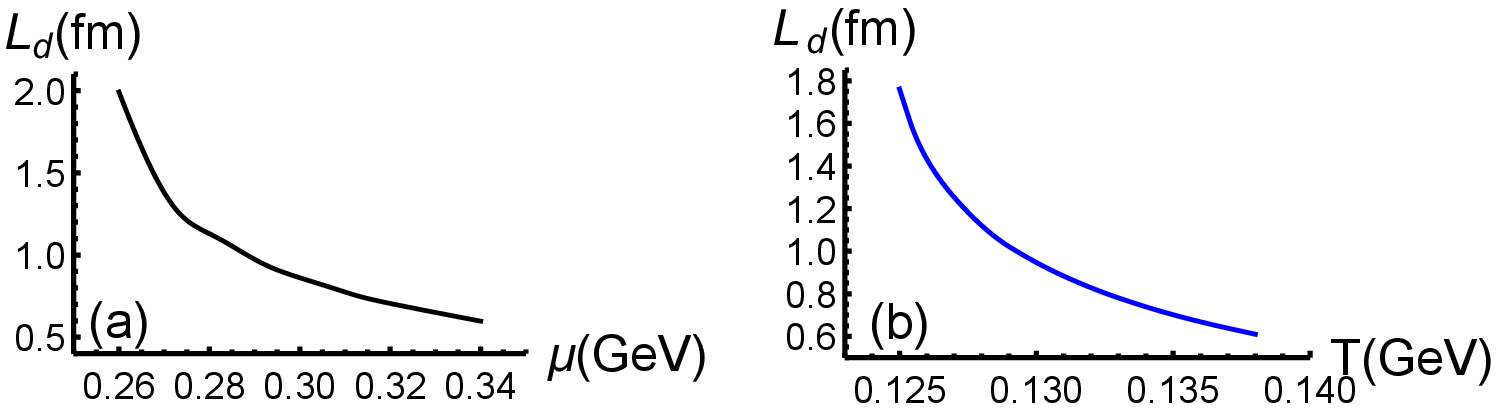}
	\caption{\label{LdMutfixed} (a) The screening distance $L_d$ as a function of $\mu$ at $T=0.1 \rm GeV$. (b) The screening distance $L_d$ as a function of $T$ at $\mu=0.1 \rm GeV$.}
\end{figure}

When string melting occurs, the diagram of $L-r_{v}$ is shown in Fig.~\ref{beyondrw}. To study the $\mu$-dependent of string melting, we fixed $T$=0.1 GeV and $\mu$ to 0.3 GeV, 0.35 GeV, and 0.4 GeV, respectively. The results of $E-L$ at different chemical potentials are presented in Fig.~\ref{melting}. As the string of $\rm{Q \bar{Q}}$ also melts under $T$=0.1 GeV and $\mu$=0.35 GeV, we compare the potential of $\rm{QQq}$ and $\rm Q\bar{Q}$ in Fig.~\ref{t01}. When $\mu$ is fixed to 0.05 GeV, we plot $E$ as a function of $L$ at different $T$ in Fig.~\ref{mufixed}. The correspondence between $E$ and $L$ is almost independent of temperature until the string melts. Furthermore, we study the melt of $\rm{QQq}$ at the different temperatures and chemical potentials in Fig.~\ref{LdMutfixed}. It is found that the influence of temperature is larger than chemical potential.

As can be seen from Fig.~\ref{melting}, with the increase of $\mu$, the melting point moves to the lower left. This means that string melting is more likely to occur at higher chemical potentials. From Fig.~\ref{Lrv0001} and Fig.~\ref{beyondrw}, we can compare the differences of $\rm{QQq}$ between small chemical potential and large chemical potential. When the string melting condition is not satisfied, the separation distance between quarks ($L$) increases monotonically as the $r_v$ increase.  Once string melting condition is satisfied, $L$ first increases and then decreases as $r_v$ increases. However, the decreasing part of the $L$ indicated by a dashed line is unphysical.

\begin{figure}
	\centering
	\includegraphics[width=8cm]{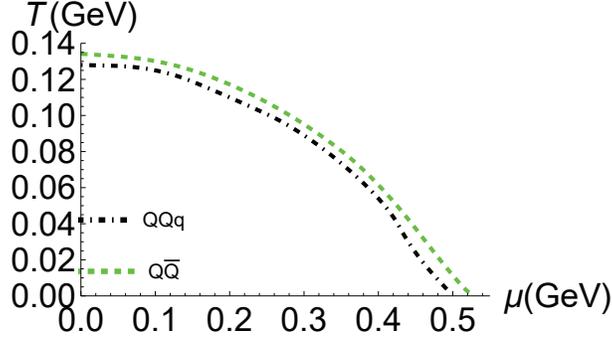}
	\caption{\label{mutplane}  The melting diagram of doubly heavy baryon and quark-antiquark pairs in the $T-\mu$ plane in units GeV.}
\end{figure}

Furthermore, the melting diagram can be drawn in a $T-\mu$ plane if enough data of $(\mu,T)$ points are found. The comparison with $\rm{Q \bar{Q}}$ \cite{Jiang:2022zbt} is also shown in Fig.~\ref{mutplane}. From the diagram above, the melting line of $\rm{QQq}$ is lower than quark-antiquark pairs.  In other words, we can infer that the quark-antiquark pair is more stable than the doubly heavy baryon.

\subsection{Two critical chemical potentials}
\begin{figure}
	\centering
	\includegraphics[width=8cm]{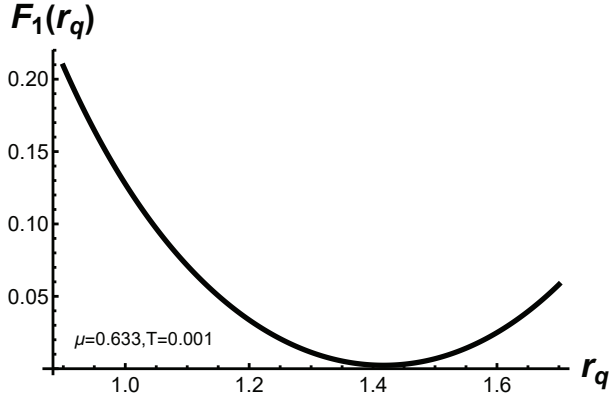}
	\caption{\label{specialvalue} The force balance equations as a function of $r_q$ in the first configuration. $r_q$ is the position of light quark, $\mu_{c2}$=0.633 GeV, and $T$=0.001 GeV. The unit of $r_q$ is $\rm{GeV^{-1}}$.}
\end{figure}
In this section, we discuss the relation between possible string configurations and $\mu$. It is found that the presence of temperature and chemical potential can make $\rm{QQq}$ unstable. By setting $T$ close to zero, we can determine the chemical potential at which Eq.~(\ref{balance equation 1}) has no solutions. We have found the critical value of the chemical potential to be 0.633 GeV at $T$=0.001 GeV, which means the first configurations cannot exist. The result is presented in Fig.~\ref{specialvalue}. At the same temperature, the $\rm{QQq}$ state melts at intermediate separation distances for $\mu$ = 0.498 GeV. The string will melt at the second configuration. We can conclude that $\rm{QQq}$ has three configurations at small chemical potential and the temperature close to zero. When $\mu$ $\textgreater$ 0.498 GeV, only two configurations can exist. Beyond 0.633 GeV, even the first configuration of $\rm{QQq}$ cannot exist.

\section{Summary and Conclusions}\label{sec:04}
In this paper, we mainly discussed the potential energy, the string breaking and $\rm{QQq}$ melting at finite temperature and chemical potential through a five-dimensional effective string model. Even with the introduction of temperature and chemical potential, the overall potential energy behavior still exhibits a Coulomb potential at small separate distances, becoming a linear potential at large separate distances. As the chemical potential increases, string breaking occurs at a larger separate distance and lower potential energy in the decay mode $\rm{Q Q q \rightarrow Q q q+Q \bar{q}}$. We discussed the string melting of $\rm{QQq}$ at a fixed temperature and chemical potential, respectively, and compared them with $\rm{Q \bar{Q}}$. It is found that the potential energy of $\rm{QQq}$ and $\rm Q\bar{Q}$ is very close when the string melts. In both cases, $\rm{QQq}$ melts at a smaller separate distance. Finally, we draw the melting diagram of $\rm{QQq}$ and $\rm{Q \bar{Q}}$ in the $T-\mu$ plane. The melting line of $\rm{QQq}$ is below that of $\rm{Q \bar{Q}}$, which indicates that $\rm{QQq}$ is less stable than ${\rm{Q\bar{Q}}}$.

\section*{Acknowledgments}
This work is supported by the Natural Science Foundation of Hunan Province of China under Grant No. 2022JJ40344, the Research Foundation of Education Bureau of Hunan Province, China (Grant No. 21B0402) and the National Natural Science Foundation of China (Grant No. 12175100).


\section*{References}

\begin{thebibliography}{120}

%\cite{Lang:1982tj}
\bibitem{Lang:1982tj}
C.~B.~Lang and C.~Rebbi,
%``Potential and Restoration of Rotational Symmetry in SU(2) Lattice Gauge Theory,''
Phys. Lett. B \textbf{115}, 137 (1982)
doi:10.1016/0370-2693(82)90813-9
%97 citations counted in INSPIRE as of 26 Apr 2023

%\cite{Hoek:1987uy}
\bibitem{Hoek:1987uy}
J.~Hoek,
%``Wilson Loops on 32**4 Lattices and the SU(3) Potential,''
Z. Phys. C \textbf{35}, 369-377 (1987)
doi:10.1007/BF01570774
%8 citations counted in INSPIRE as of 26 Apr 2023

%\cite{Michael:1990az}
\bibitem{Michael:1990az}
C.~Michael and S.~J.~Perantonis,
%``Potentials and glueballs at large beta in SU(2) pure gauge theory,''
J. Phys. G \textbf{18}, 1725-1736 (1992)
doi:10.1088/0954-3899/18/11/005
%26 citations counted in INSPIRE as of 26 Apr 2023

%\cite{Takahashi:2002bw}
\bibitem{Takahashi:2002bw}
T.~T.~Takahashi, H.~Suganuma, Y.~Nemoto and H.~Matsufuru,
%``Detailed analysis of the three quark potential in SU(3) lattice QCD,''
Phys. Rev. D \textbf{65}, 114509 (2002)
doi:10.1103/PhysRevD.65.114509
[arXiv:hep-lat/0204011 [hep-lat]].
%251 citations counted in INSPIRE as of 26 Apr 2023

%\cite{Aoki:2005vt}
\bibitem{Aoki:2005vt}
Y.~Aoki, Z.~Fodor, S.~D.~Katz and K.~K.~Szabo,
%``The Equation of state in lattice QCD: With physical quark masses towards the continuum limit,''
JHEP \textbf{01}, 089 (2006)
doi:10.1088/1126-6708/2006/01/089
[arXiv:hep-lat/0510084 [hep-lat]].
%338 citations counted in INSPIRE as of 26 Apr 2023

%\cite{Ratti:2005jh}
\bibitem{Ratti:2005jh}
C.~Ratti, M.~A.~Thaler and W.~Weise,
%``Phases of QCD: Lattice thermodynamics and a field theoretical model,''
Phys. Rev. D \textbf{73}, 014019 (2006)
doi:10.1103/PhysRevD.73.014019
[arXiv:hep-ph/0506234 [hep-ph]].
%905 citations counted in INSPIRE as of 26 Apr 2023

%\cite{Bicudo:2007xp}
\bibitem{Bicudo:2007xp}
P.~Bicudo, M.~Cardoso and O.~Oliveira,
%``Study of the gluon-quark-antiquark static potential in SU(3) lattice QCD,''
Phys. Rev. D \textbf{77}, 091504 (2008)
doi:10.1103/PhysRevD.77.091504
[arXiv:0704.2156 [hep-lat]].
%28 citations counted in INSPIRE as of 26 Apr 2023

%\cite{Luscher:2010iy}
\bibitem{Luscher:2010iy}
M.~L\"uscher,
%``Properties and uses of the Wilson flow in lattice QCD,''
JHEP \textbf{08}, 071 (2010)
[erratum: JHEP \textbf{03}, 92 (2014)]
doi:10.1007/JHEP08(2010)071
[arXiv:1006.4518 [hep-lat]].

%\cite{Hasenfratz:1983ba}
\bibitem{Hasenfratz:1983ba}
P.~Hasenfratz and F.~Karsch,
%``Chemical Potential on the Lattice,''
Phys. Lett. B \textbf{125}, 308-310 (1983)
doi:10.1016/0370-2693(83)91290-X
%410 citations counted in INSPIRE as of 23 May 2023

%\cite{Fodor:2001au}
\bibitem{Fodor:2001au}
Z.~Fodor and S.~D.~Katz,
%``A New method to study lattice QCD at finite temperature and chemical potential,''
Phys. Lett. B \textbf{534}, 87-92 (2002)
doi:10.1016/S0370-2693(02)01583-6
[arXiv:hep-lat/0104001 [hep-lat]].
%598 citations counted in INSPIRE as of 26 Apr 2023

%\cite{Muroya:2003qs}
\bibitem{Muroya:2003qs}
S.~Muroya, A.~Nakamura, C.~Nonaka and T.~Takaishi,
%``Lattice QCD at finite density: An Introductory review,''
Prog. Theor. Phys. \textbf{110}, 615-668 (2003)
doi:10.1143/PTP.110.615
[arXiv:hep-lat/0306031 [hep-lat]].
%215 citations counted in INSPIRE as of 26 Apr 2023
%929 citations counted in INSPIRE as of 26 Apr 2023

%\cite{Brambilla:2006zx}
\bibitem{Brambilla:2006zx}
N.~Brambilla,
%``Systems of two heavy quarks with effective field theories,''
doi:10.1142/9789812708267\_0014
[arXiv:hep-ph/0609237 [hep-ph]].
%2 citations counted in INSPIRE as of 23 May 2023

%\cite{Ghiglieri:2011fhu}
\bibitem{Ghiglieri:2011fhu}
J.~Ghiglieri,
%``Effective Field Theories of QCD for Heavy Quarkonia at Finite Temperature,''
[arXiv:1201.2920 [hep-ph]].
%12 citations counted in INSPIRE as of 23 May 2023

	%\cite{Andreev:2020pqy}
\bibitem{Andreev:2020pqy}
O.~Andreev,
%``String Breaking, Baryons, Medium, and Gauge/String Duality,''
Phys. Rev. D \textbf{101}, no.10, 106003 (2020)
doi:10.1103/PhysRevD.101.106003
[arXiv:2003.09880 [hep-ph]].
%4 citations counted in INSPIRE as of 28 Aug 2021

%\cite{Andreev:2021bfg}
\bibitem{Andreev:2021bfg}
O.~Andreev,
%``Remarks on static three-quark potentials, string breaking and gauge/string duality,''
Phys. Rev. D \textbf{104}, no.2, 026005 (2021)
doi:10.1103/PhysRevD.104.026005
[arXiv:2101.03858 [hep-ph]].
%1 citations counted in INSPIRE as of 29 Aug 2021

%\cite{Andreev:2019cbc}
\bibitem{Andreev:2019cbc}
O.~Andreev,
%``Baryon modes in string breaking from gauge/string duality,''
Phys. Lett. B \textbf{804}, 135406 (2020)
doi:10.1016/j.physletb.2020.135406
[arXiv:1909.12771 [hep-ph]].
%11 citations counted in INSPIRE as of 12 Apr 2023

%\cite{Eichten:1978tg}
\bibitem{Eichten:1978tg}
E.~Eichten, K.~Gottfried, T.~Kinoshita, K.~D.~Lane and T.~M.~Yan,
%``Charmonium: The Model,''
Phys. Rev. D \textbf{17}, 3090 (1978)
[erratum: Phys. Rev. D \textbf{21}, 313 (1980)]
doi:10.1103/PhysRevD.17.3090
%1684 citations counted in INSPIRE as of 12 Apr 2023
	
%\cite{Eichten:1979ms}
\bibitem{Eichten:1979ms}
E.~Eichten, K.~Gottfried, T.~Kinoshita, K.~D.~Lane and T.~M.~Yan,
%``Charmonium: Comparison with Experiment,''
Phys. Rev. D \textbf{21}, 203 (1980)
doi:10.1103/PhysRevD.21.203
%1880 citations counted in INSPIRE as of 12 Apr 2023	
	
	%\cite{Andreev:2020xor}
\bibitem{Andreev:2020xor}
O.~Andreev,
%``Some Properties of the $QQq$-Quark Potential in String Models,''
JHEP \textbf{05}, 173 (2021)
doi:10.1007/JHEP05(2021)173
[arXiv:2007.15466 [hep-ph]].
%7 citations counted in INSPIRE as of 12 Apr 2023



%\cite{Sumino:2004ht}
\bibitem{Sumino:2004ht}
Y.~Sumino,
%``'Coulomb + linear' form of the static QCD potential in operator product expansion,''
Phys. Lett. B \textbf{595}, 387-392 (2004)
doi:10.1016/j.physletb.2004.06.065
[arXiv:hep-ph/0403242 [hep-ph]].
%17 citations counted in INSPIRE as of 21 Apr 2023

%\cite{Andreev:2006eh}
\bibitem{Andreev:2006eh}
O.~Andreev and V.~I.~Zakharov,
%``The Spatial String Tension, Thermal Phase Transition, and AdS/QCD,''
Phys. Lett. B \textbf{645}, 437-441 (2007)
doi:10.1016/j.physletb.2007.01.002
[arXiv:hep-ph/0607026 [hep-ph]].
%116 citations counted in INSPIRE as of 21 Apr 2023

%\cite{Balitsky:1985iw}
\bibitem{Balitsky:1985iw}
I.~I.~Balitsky,
%``WILSON LOOP FOR THE STRETCHED CONTOURS IN VACUUM FIELDS AND THE SMALL DISTANCE BEHAVIOR OF THE INTERQUARK POTENTIAL,''
Nucl. Phys. B \textbf{254}, 166-186 (1985)
doi:10.1016/0550-3213(85)90215-9
%111 citations counted in INSPIRE as of 21 Apr 2023

%\cite{deForcrand:2005vv}
\bibitem{deForcrand:2005vv}
P.~de Forcrand and O.~Jahn,
%``The Baryon static potential from lattice QCD,''
Nucl. Phys. A \textbf{755}, 475-480 (2005)
doi:10.1016/j.nuclphysa.2005.03.127
[arXiv:hep-ph/0502039 [hep-ph]].
%35 citations counted in INSPIRE as of 21 Apr 2023

%\cite{Andreev:2006nw}
\bibitem{Andreev:2006nw}
O.~Andreev and V.~I.~Zakharov,
%``On Heavy-Quark Free Energies, Entropies, Polyakov Loop, and AdS/QCD,''
JHEP \textbf{04}, 100 (2007)
doi:10.1088/1126-6708/2007/04/100
[arXiv:hep-ph/0611304 [hep-ph]].
%64 citations counted in INSPIRE as of 24 Apr 2023

%\cite{He:2010bx}
\bibitem{He:2010bx}
S.~He, M.~Huang and Q.~s.~Yan,
%``Heavy quark potential and QCD beta function from a deformed $AdS_5$ model,''
Prog. Theor. Phys. Suppl. \textbf{186}, 504-509 (2010)
doi:10.1143/PTPS.186.504
[arXiv:1007.0088 [hep-ph]].
%8 citations counted in INSPIRE as of 24 Apr 2023

%\cite{Colangelo:2010pe}
\bibitem{Colangelo:2010pe}
P.~Colangelo, F.~Giannuzzi and S.~Nicotri,
%``Holography, Heavy-Quark Free Energy, and the QCD Phase Diagram,''
Phys. Rev. D \textbf{83}, 035015 (2011)
doi:10.1103/PhysRevD.83.035015
[arXiv:1008.3116 [hep-ph]].
%74 citations counted in INSPIRE as of 24 Apr 2023

%\cite{DeWolfe:2010he}
\bibitem{DeWolfe:2010he}
O.~DeWolfe, S.~S.~Gubser and C.~Rosen,
%``A holographic critical point,''
Phys. Rev. D \textbf{83}, 086005 (2011)
doi:10.1103/PhysRevD.83.086005
[arXiv:1012.1864 [hep-th]].
%158 citations counted in INSPIRE as of 24 Apr 2023


%\cite{Li:2011hp}
\bibitem{Li:2011hp}
D.~Li, S.~He, M.~Huang and Q.~S.~Yan,
%``Thermodynamics of deformed AdS$_5$ model with a positive/negative quadratic correction in graviton-dilaton system,''
JHEP \textbf{09}, 041 (2011)
doi:10.1007/JHEP09(2011)041
[arXiv:1103.5389 [hep-th]].
%92 citations counted in INSPIRE as of 24 Apr 2023

%\cite{Fadafan:2011gm}
\bibitem{Fadafan:2011gm}
K.~B.~Fadafan,
%``Heavy quarks in the presence of higher derivative corrections from AdS/CFT,''
Eur. Phys. J. C \textbf{71}, 1799 (2011)
doi:10.1140/epjc/s10052-011-1799-7
[arXiv:1102.2289 [hep-th]].
%27 citations counted in INSPIRE as of 24 Apr 2023

%\cite{Fadafan:2012qy}
\bibitem{Fadafan:2012qy}
K.~B.~Fadafan and E.~Azimfard,
%``On meson melting in the quark medium,''
Nucl. Phys. B \textbf{863}, 347-360 (2012)
doi:10.1016/j.nuclphysb.2012.05.022
[arXiv:1203.3942 [hep-th]].
%19 citations counted in INSPIRE as of 24 Apr 2023

%\cite{Cai:2012xh}
\bibitem{Cai:2012xh}
R.~G.~Cai, S.~He and D.~Li,
%``A hQCD model and its phase diagram in Einstein-Maxwell-Dilaton system,''
JHEP \textbf{03}, 033 (2012)
doi:10.1007/JHEP03(2012)033
[arXiv:1201.0820 [hep-th]].
%76 citations counted in INSPIRE as of 24 Apr 2023

%\cite{Li:2012ay}
\bibitem{Li:2012ay}
D.~Li, M.~Huang and Q.~S.~Yan,
%``A dynamical soft-wall holographic QCD model for chiral symmetry breaking and linear confinement,''
Eur. Phys. J. C \textbf{73}, 2615 (2013)
doi:10.1140/epjc/s10052-013-2615-3
[arXiv:1206.2824 [hep-th]].
%63 citations counted in INSPIRE as of 24 Apr 2023

%\cite{Fang:2015ytf}
\bibitem{Fang:2015ytf}
Z.~Fang, S.~He and D.~Li,
%``Chiral and Deconfining Phase Transitions from Holographic QCD Study,''
Nucl. Phys. B \textbf{907}, 187-207 (2016)
doi:10.1016/j.nuclphysb.2016.04.003
[arXiv:1512.04062 [hep-ph]].
%35 citations counted in INSPIRE as of 24 Apr 2023

%\cite{Yang:2015aia}
\bibitem{Yang:2015aia}
Y.~Yang and P.~H.~Yuan,
%``Confinement-deconfinement phase transition for heavy quarks in a soft wall holographic QCD model,''
JHEP \textbf{12}, 161 (2015)
doi:10.1007/JHEP12(2015)161
[arXiv:1506.05930 [hep-th]].
%58 citations counted in INSPIRE as of 24 Apr 2023

%\cite{Zhang:2015faa}
\bibitem{Zhang:2015faa}
Z.~q.~Zhang, D.~f.~Hou and G.~Chen,
%``Heavy quark potential from deformed $AdS_5$ models,''
Nucl. Phys. A \textbf{960}, 1-10 (2017)
doi:10.1016/j.nuclphysa.2017.01.007
[arXiv:1507.07263 [hep-ph]].
%5 citations counted in INSPIRE as of 24 Apr 2023

%\cite{Ewerz:2016zsx}
\bibitem{Ewerz:2016zsx}
C.~Ewerz, O.~Kaczmarek and A.~Samberg,
%``Free Energy of a Heavy Quark-Antiquark Pair in a Thermal Medium from AdS/CFT,''
JHEP \textbf{03}, 088 (2018)
doi:10.1007/JHEP03(2018)088
[arXiv:1605.07181 [hep-th]].
%21 citations counted in INSPIRE as of 24 Apr 2023

%\cite{Chen:2017lsf}
\bibitem{Chen:2017lsf}
X.~Chen, S.~Q.~Feng, Y.~F.~Shi and Y.~Zhong,
%``Moving heavy quarkonium entropy, effective string tension, and the QCD phase diagram,''
Phys. Rev. D \textbf{97}, no.6, 066015 (2018)
doi:10.1103/PhysRevD.97.066015
[arXiv:1710.00465 [hep-ph]].
%18 citations counted in INSPIRE as of 24 Apr 2023

%\cite{Arefeva:2018hyo}
\bibitem{Arefeva:2018hyo}
I.~Aref'eva and K.~Rannu,
%``Holographic Anisotropic Background with Confinement-Deconfinement Phase Transition,''
JHEP \textbf{05}, 206 (2018)
doi:10.1007/JHEP05(2018)206
[arXiv:1802.05652 [hep-th]].
%69 citations counted in INSPIRE as of 24 Apr 2023

%\cite{Chen:2018vty}
\bibitem{Chen:2018vty}
X.~Chen, D.~Li and M.~Huang,
%``Criticality of QCD in a holographic QCD model with critical end point,''
Chin. Phys. C \textbf{43}, no.2, 023105 (2019)
doi:10.1088/1674-1137/43/2/023105
[arXiv:1810.02136 [hep-ph]].
%20 citations counted in INSPIRE as of 24 Apr 2023

%\cite{Bohra:2019ebj}
\bibitem{Bohra:2019ebj}
H.~Bohra, D.~Dudal, A.~Hajilou and S.~Mahapatra,
%``Anisotropic string tensions and inversely magnetic catalyzed deconfinement from a dynamical AdS/QCD model,''
Phys. Lett. B \textbf{801}, 135184 (2020)
doi:10.1016/j.physletb.2019.135184
[arXiv:1907.01852 [hep-th]].
%50 citations counted in INSPIRE as of 24 Apr 2023

%\cite{Chen:2019rez}
\bibitem{Chen:2019rez}
X.~Chen, D.~Li, D.~Hou and M.~Huang,
%``Quarkyonic phase from quenched dynamical holographic QCD model,''
JHEP \textbf{03}, 073 (2020)
doi:10.1007/JHEP03(2020)073
[arXiv:1908.02000 [hep-ph]].
%29 citations counted in INSPIRE as of 24 Apr 2023

%\cite{Zhou:2020ssi}
\bibitem{Zhou:2020ssi}
J.~Zhou, X.~Chen, Y.~Q.~Zhao and J.~Ping,
%``Thermodynamics of heavy quarkonium in a magnetic field background,''
Phys. Rev. D \textbf{102}, no.8, 086020 (2020)
doi:10.1103/PhysRevD.102.086020
[arXiv:2006.09062 [hep-ph]].
%15 citations counted in INSPIRE as of 24 Apr 2023

%\cite{Zhou:2021sdy}
\bibitem{Zhou:2021sdy}
J.~Zhou, X.~Chen, Y.~Q.~Zhao and J.~Ping,
%``Thermodynamics of heavy quarkonium in rotating matter from holography,''
Phys. Rev. D \textbf{102}, no.12, 126029 (2021)
doi:10.1103/PhysRevD.102.126029
%10 citations counted in INSPIRE as of 24 Apr 2023


%\cite{Chen:2020ath}
\bibitem{Chen:2020ath}
X.~Chen, L.~Zhang, D.~Li, D.~Hou and M.~Huang,
%``Gluodynamics and deconfinement phase transition under rotation from holography,''
JHEP \textbf{07}, 132 (2021)
doi:10.1007/JHEP07(2021)132
[arXiv:2010.14478 [hep-ph]].
%37 citations counted in INSPIRE as of 24 Apr 2023

%\cite{Chen:2021gop}
\bibitem{Chen:2021gop}
X.~Chen, L.~Zhang and D.~Hou,
%``Running coupling constant at finite chemical potential and magnetic field from holography *,''
Chin. Phys. C \textbf{46}, no.7, 073101 (2022)
doi:10.1088/1674-1137/ac5c2d
[arXiv:2108.03840 [hep-ph]].
%4 citations counted in INSPIRE as of 24 Apr 2023

%\cite{Gross:1998gk}
\bibitem{Gross:1998gk}
D.~J.~Gross and H.~Ooguri,
%``Aspects of large N gauge theory dynamics as seen by string theory,''
Phys. Rev. D \textbf{58}, 106002 (1998)
doi:10.1103/PhysRevD.58.106002
[arXiv:hep-th/9805129 [hep-th]].
%402 citations counted in INSPIRE as of 12 Apr 2023
%\cite{Andreev:2006nw,He:2010bx,Colangelo:2010pe,DeWolfe:2010he,Li:2011hp,Fadafan:2011gm,Fadafan:2012qy,Cai:2012xh,Li:2012ay,Fang:2015ytf,Yang:2015aia,Zhang:2015faa,Ewerz:2016zsx,Chen:2017lsf,Arefeva:2018hyo,Chen:2018vty,Bohra:2019ebj,Bohra:2019ebj,Chen:2019rez,Zhou:2020ssi,{Zhou:2021sdy,Chen:2020ath,Chen:2021gop,Gross:1998gk}
%4-27结束




%\cite{Maldacena:1997re}
\bibitem{Maldacena:1997re}
J.~M.~Maldacena,
%``The Large N limit of superconformal field theories and supergravity,''
Adv. Theor. Math. Phys. \textbf{2}, 231-252 (1998)
doi:10.4310/ATMP.1998.v2.n2.a1
[arXiv:hep-th/9711200 [hep-th]].
%18457 citations counted in INSPIRE as of 12 Apr 2023

%\cite{Maldacena:1998im}
\bibitem{Maldacena:1998im}
J.~M.~Maldacena,
%``Wilson loops in large N field theories,''
Phys. Rev. Lett. \textbf{80}, 4859-4862 (1998)
doi:10.1103/PhysRevLett.80.4859
[arXiv:hep-th/9803002 [hep-th]].
%1908 citations counted in INSPIRE as of 12 Apr 2023

%\cite{Rey:1998ik}
\bibitem{Rey:1998ik}
S.~J.~Rey and J.~T.~Yee,
%``Macroscopic strings as heavy quarks in large N gauge theory and anti-de Sitter supergravity,''
Eur. Phys. J. C \textbf{22}, 379-394 (2001)
doi:10.1007/s100520100799
[arXiv:hep-th/9803001 [hep-th]].
%1412 citations counted in INSPIRE as of 12 Apr 2023

%\cite{Rey:1998bq}
\bibitem{Rey:1998bq}
S.~J.~Rey, S.~Theisen and J.~T.~Yee,
%``Wilson-Polyakov loop at finite temperature in large N gauge theory and anti-de Sitter supergravity,''
Nucl. Phys. B \textbf{527}, 171-186 (1998)
doi:10.1016/S0550-3213(98)00471-4
[arXiv:hep-th/9803135 [hep-th]].
%458 citations counted in INSPIRE as of 12 Apr 2023

%\cite{Witten:1998xy}
\bibitem{Witten:1998xy}
E.~Witten,
%``Baryons and branes in anti-de Sitter space,''
JHEP \textbf{07}, 006 (1998)
doi:10.1088/1126-6708/1998/07/006
[arXiv:hep-th/9805112 [hep-th]].
%722 citations counted in INSPIRE as of 12 Apr 2023

%\cite{Andreev:2015iaa}
\bibitem{Andreev:2015iaa}
O.~Andreev,
%``Model of the $N$-quark potential in $SU(N)$ gauge theory using gauge-string duality,''
Phys. Lett. B \textbf{756}, 6-9 (2016)
doi:10.1016/j.physletb.2016.02.070
[arXiv:1505.01067 [hep-ph]].
%13 citations counted in INSPIRE as of 12 Apr 2023

%\cite{Andreev:2015riv}
\bibitem{Andreev:2015riv}
O.~Andreev,
%``Some Aspects of Three-Quark Potentials,''
Phys. Rev. D \textbf{93}, no.10, 105014 (2016)
doi:10.1103/PhysRevD.93.105014
[arXiv:1511.03484 [hep-ph]].
%19 citations counted in INSPIRE as of 12 Apr 2023


%\cite{Andreev:2006ct}
\bibitem{Andreev:2006ct}
O.~Andreev and V.~I.~Zakharov,
%``Heavy-quark potentials and AdS/QCD,''
Phys. Rev. D \textbf{74}, 025023 (2006)
doi:10.1103/PhysRevD.74.025023
[arXiv:hep-ph/0604204 [hep-ph]].
%272 citations counted in INSPIRE as of 12 Apr 2023

%\cite{Andreev:2012mc}
\bibitem{Andreev:2012mc}
O.~Andreev,
%``Exotic Hybrid Quark Potentials,''
Phys. Rev. D \textbf{86}, 065013 (2012)
doi:10.1103/PhysRevD.86.065013
[arXiv:1207.1892 [hep-ph]].
%14 citations counted in INSPIRE as of 12 Apr 2023



%\cite{Xun chen:2022bq}
\bibitem{Xun chen:2022bq}
Xun Chen, Bo Yu, Peng-Cheng Chu, Xiao-hua Li,
Chin.Phys.C 46 (2022) 7, 073102
doi:10.1088/1674-1137/ac5db9
[arXiv:2112.06234 [hep-ph]].
%445 citations counted in INSPIRE as of 29 Aug 2021


%\cite{Jiang:2022zbt}
\bibitem{Jiang:2022zbt}
J.~J.~Jiang, Y.~Z.~Xiao, J.~Qin, X.~Li and X.~Chen,
%``Three-quark potential at finite temperature and chemical potential*,''
Chin. Phys. C \textbf{47}, no.1, 013106 (2023)
doi:10.1088/1674-1137/ac9894
[arXiv:2212.03541 [hep-ph]].
%0 citations counted in INSPIRE as of 12 Apr 2023		
	
\end{thebibliography}
\end{document}